\documentclass[a4paper,11pt]{article}
\pdfoutput=1 

\usepackage{jinstpub} 

\title{A GEM based TPC for beam monitoring}

\author[a,b,1]{G. Galg\'oczi,\note{Corresponding author.}}
\author[a]{G. Hamar,}
\author[a]{P. P\'azm\'andi,}
\author[a]{and D. Varga}

\affiliation[a]{Wigner Research Centre for Physics,\\1121 Budapest, Konkoly-Thege Mikl\'os\'ut 29-33., Budapest, Hungary}
\affiliation[b]{Eotvos Lor\'and University, Faculty of Sciences, Department of Physics,\\ 
Egyetem t\'er 1-3, 1053, Budapest, Hungary
 }

\emailAdd{galgoczi@caesar.elte.hu}

\abstract{
In recent years Gas Electron Multipliers \cite{1} have proven to be reliable amplification stages at high beam rates, and can be used also in Time Projection Chambers \cite{1p5}. Our group developed a 1 dm$^3$ active volume double-GEM TPC,
with spatial resolution of 50 $\mu m$ and 280 $\mu$m. Custom designed FPGA data acquisition system enables rate capability for about 100 kHz$\cdot$mm$^{-2}$, providing excellent track-by-track position and angular information, better than 0.1 mm and 1 mrad respectively. The wide dynamic range of the system enables identification from $^{4}$He up to $^{86}$Kr using ionization measurement. Two of these TPCs are planned to operate in tandem mode \cite{2,3} to filter off-time particles and to achieve a superior angular resolution.}

\keywords{beam-line instrumentation, micropattern gaseous detectors, Time Projection Chambers}

\proceeding{International Conference on Instrumentation for Colliding Beam Physics\\
  24-28 February 2020\\
  Novosibirsk, Russia}

\begin{document}
\maketitle
\flushbottom

\section{Introduction}

We present the test results of a newly developed Time Projection Chamber (TPC) which has a gas amplification stage of a double Gas Electron Multipliers (GEMs) \cite{1}. The idea is to exploit the advantages of GEMs for gaining high granularity with an outstanding ion feedback suppression. Several such detector systems are currently being developed and tested for several applications including beam monitoring \cite{beam_mon, tracker} hadron therapy \cite{hadron_therapy} and particle identification. In our case a set of commercially available FPGAs (Zynq series) with a custom designed front end electronics are capable of reading out the electronic signal with 5 ns time resolution. The results of the first test measurements conducted at RIBLL (the radioactive ion beam line at the Institute of Modern Physics, in Lanzhou, China) \cite{4} are presented and a spatial resolution of 50 $\mu$m along the pads and 280 $\mu$m (for a drift velocity of 1.3 cm/$\mu$s) in the time "direction" is derived.

\section{The detector and the read-out electronics}

The active volume of the TPC is 10x10x10 cm$^{3}$. Charge amplification is done by the two standard 10x10 cm$^{2}$ GEMs. Each GEM has a hole diameter of 70 $\mu$m, thickness of 50 $\mu$m and a pitch of 140 $\mu$m. Five segmented pad rows with a spacing of 20 mm collect the charge after the initial amplification. Each pad row consists of 64 pads. The 32 pads in the middle of the row have a size of 1.2x20 mm$^{2}$ and the outer 32 have a size of 2x20 mm$^{2}$. Finer resolution is needed in the inner pads as the beam crosses the detector there. In the outer areas we expect much lower particle multiplicity as only the scattered particles cross there. The beam enters the TPC through a 15 $\mu$m mylar window and the field cage twice (50 $\mu$m kapton and 10 $\mu$m copper). The beam-detector interactions are minimized with such a thin window and field cage. To ensure the high purity of the gas inside the active volume the TPC has a double wall system: field cage and gas cage. The high purity gas enters into the active volume from the cathode,
flows towards the GEMs, closed sideways via the gas-tight field cage; then enters in between the two cage walls before the outlet. The drift field of the TPC and the amplification gain of the GEMs are set through a resistor chain. 

\begin{figure}[h!]
\centering 
\includegraphics[width=.45\textwidth,clip]{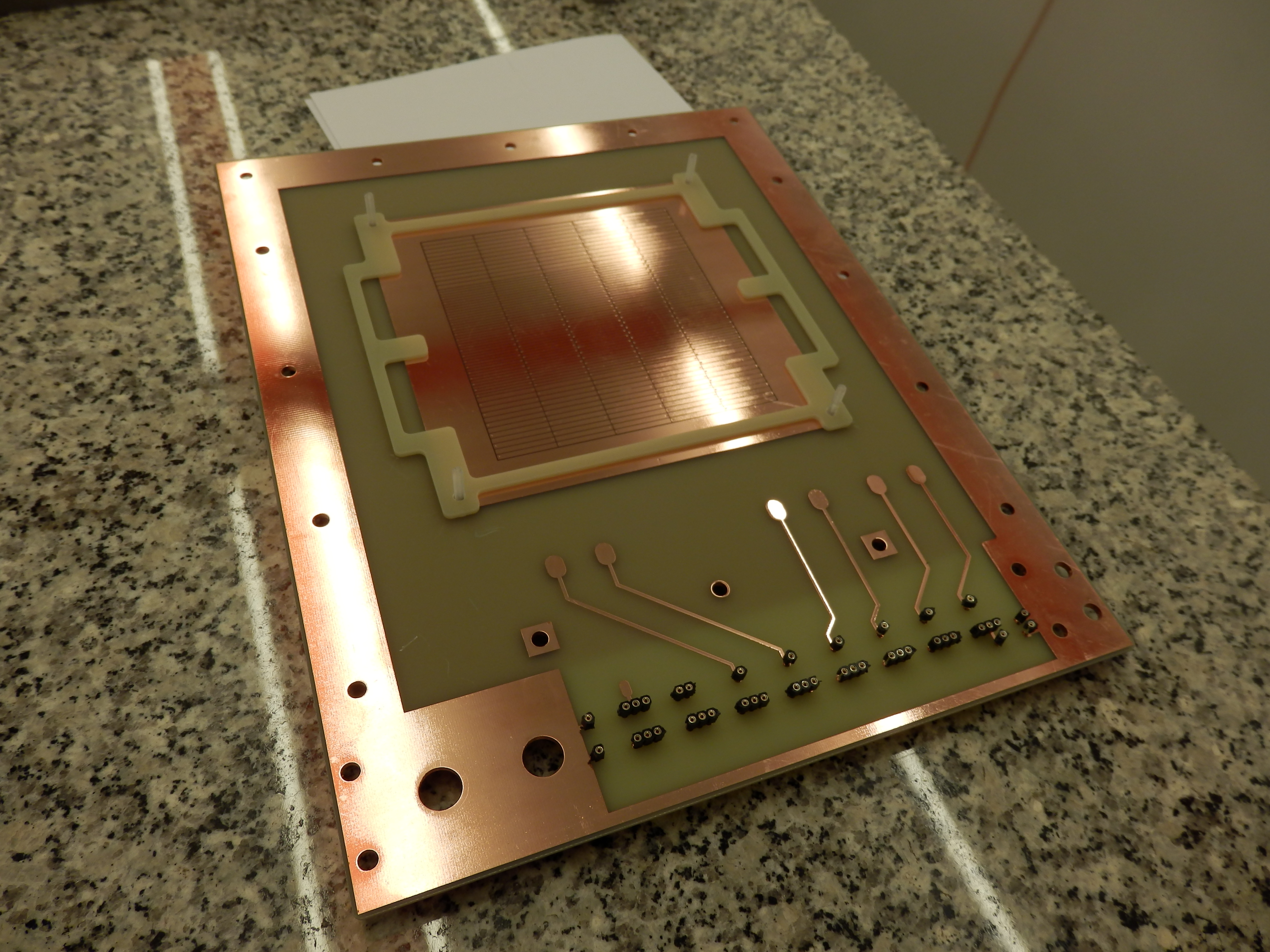}
\qquad
\includegraphics[width=.45\textwidth,origin=c]{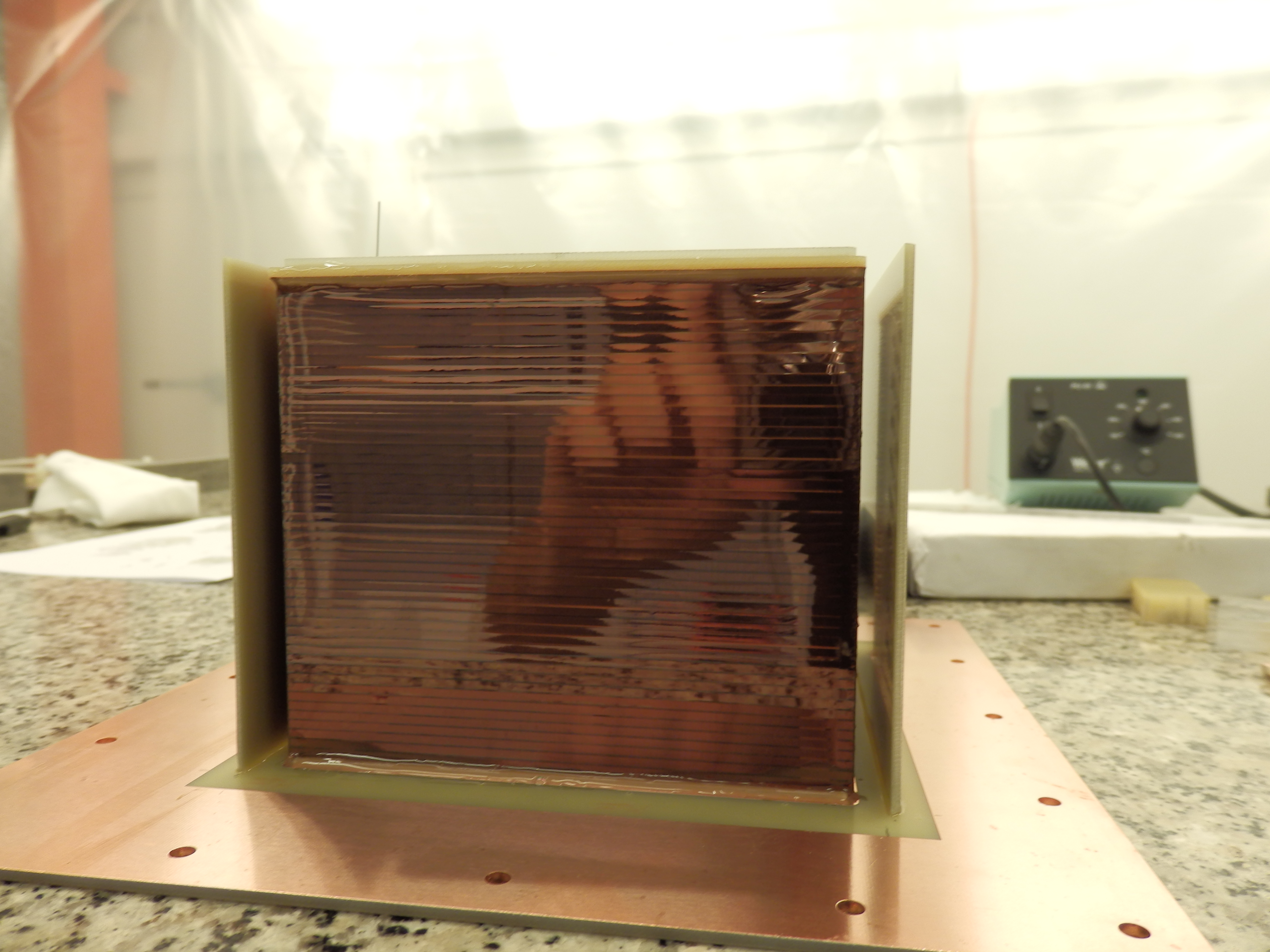}
\caption{\label{fig:design}(Left) The PadPlane of the TPC, containing the 5 padrows,
the HV connections points of the GEMs lead to the place of the resistor 
chain. (Right) The kapton field cage of the active volume, and 2mm away the 
support walls of the thin outer gas cage.}
\end{figure}

The signal from each pad is amplified by a preamplifier on the FEE card and then digitized with a 12 bit ADC at a 2.5 MHz sampling rate. The ADCs are read out with five Zynq-7020 FPGAs with a time resolution of 5 ns in order to achieve a good spatial resolution perpendicular to the pad plane.

\begin{figure}[h!]
\centering 
\includegraphics[width=.45\textwidth,clip]{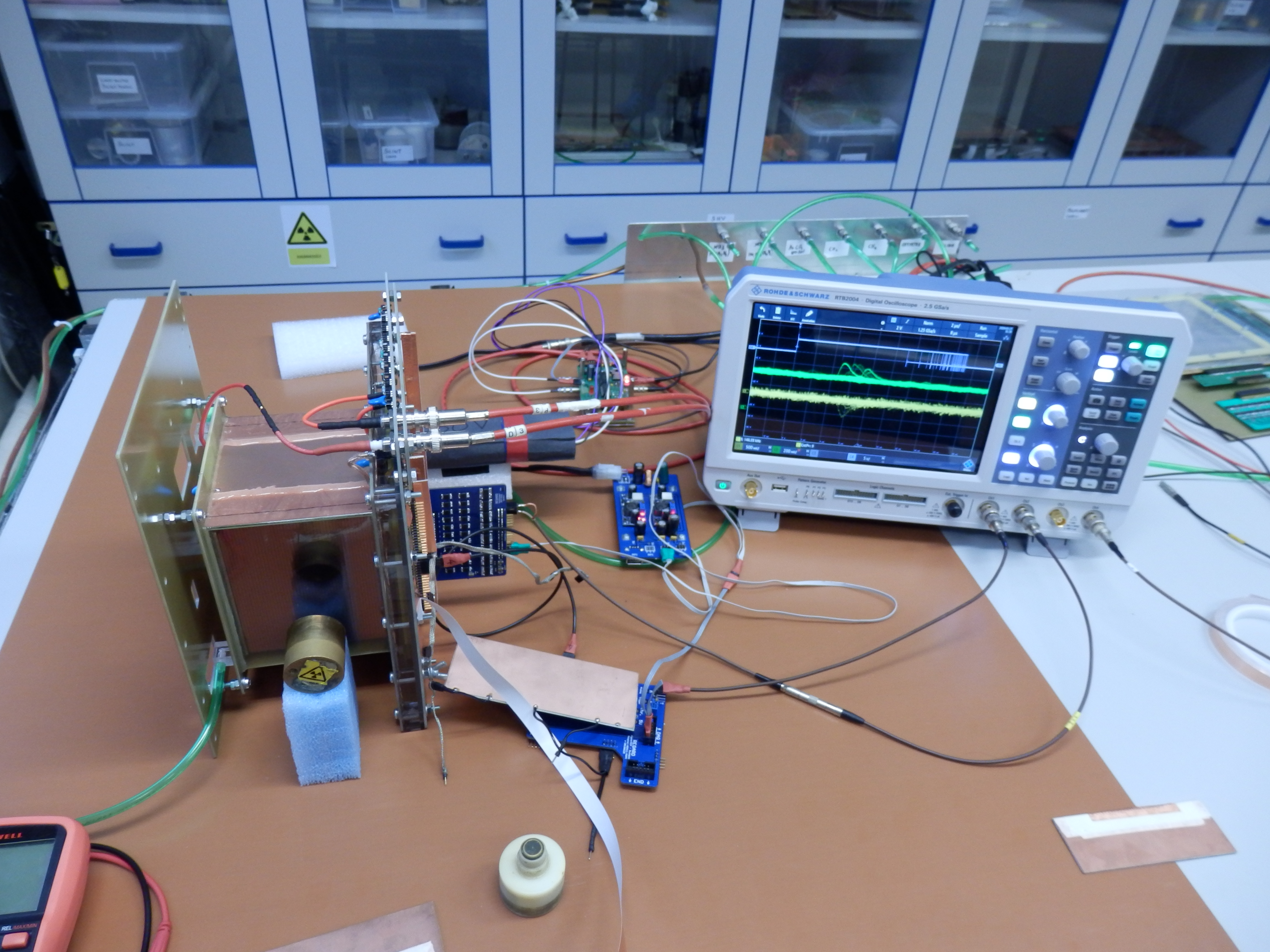}
\qquad
\includegraphics[width=.45\textwidth,origin=c]{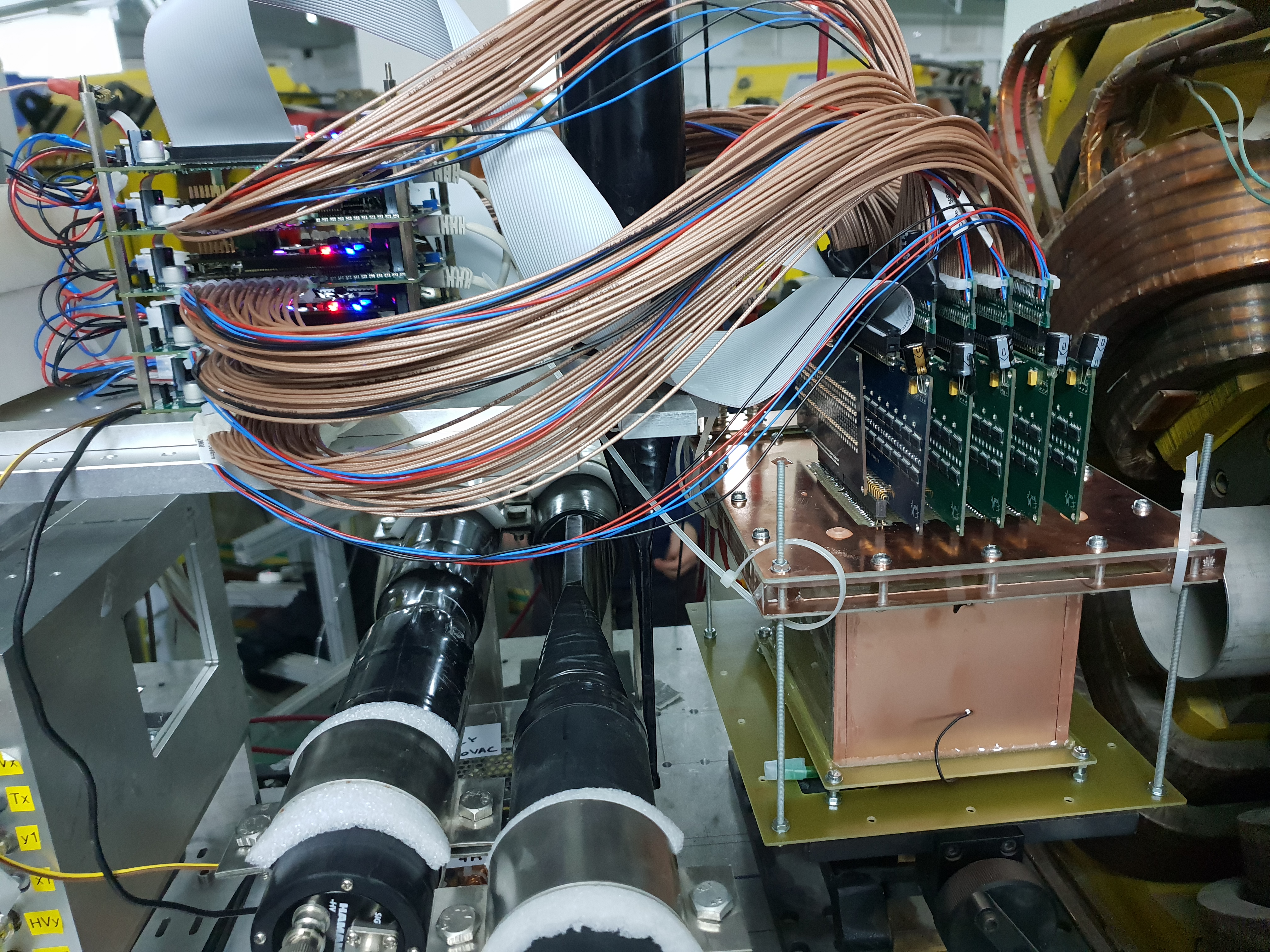}
\caption{\label{fig:azimuthal} The TPC during lab testing with a $\beta$ source (left). GEM-TPC test measurement in the RIBLL beam in Lanzhou (right). The FEE electronics with one preamplifier per channel connected to the FPGA board tower are visible.}
\end{figure}

\section{Test measurements at the radioactive ion beam line in Lanzhou}

The performance of the detector and the read-out was tested at the RIBLL \cite{4} during measurements in 2017 and 2019. An Ar/CO$_{2}$ mixture (85\% / 15\% respectively) was used in the TPC. The beam consisted of $^{86}$Kr$^{36+}$ ions, varying between 10,000 and 100,000 in each spill. Since these ions create a considerably large primary signal low GEM amplification was required. The beamspot was 3x3 mm$^2$ on the detector. The drift velocity was varied in the range of 0.2-1.5 cm/$\mu$s.

\begin{figure}[h!]
\centering 
\includegraphics[width=.45\textwidth,clip]{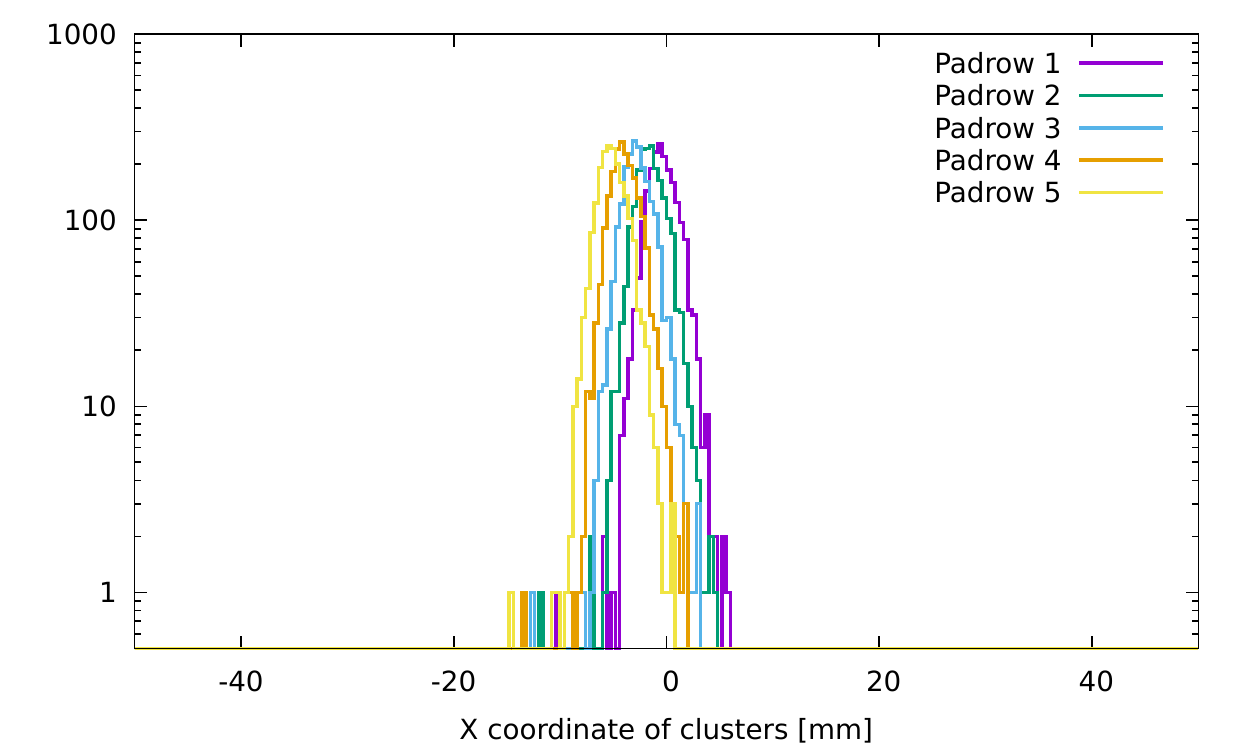}
\qquad
\includegraphics[width=.45\textwidth,origin=c]{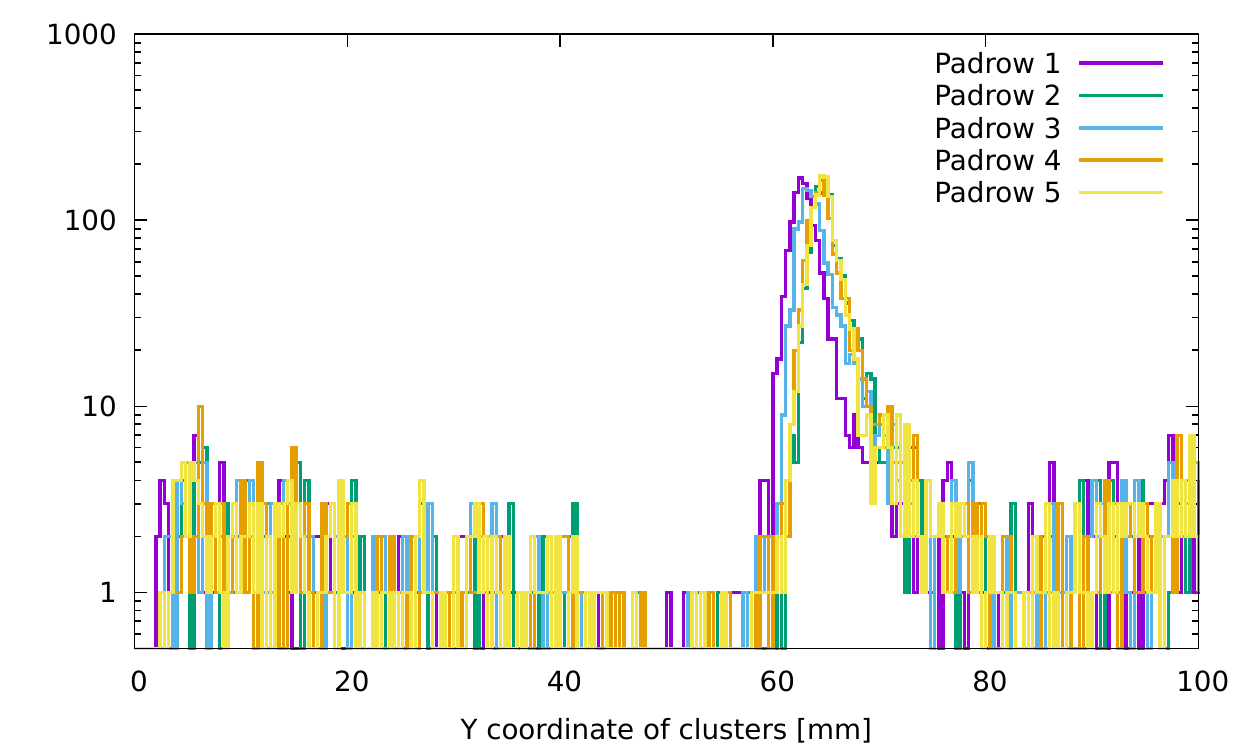}
\caption{\label{fig:center} (Left) The distribution of the center of clusters parallel to the pads, for each pad row independently. The detector was rotated by 100 mrad thus the shift of the clusters in consequent layers. (Right) Figure distribution of the center of clusters in the vertical (time) direction, for each pad row independently. Some off-time ions are observable.}
\end{figure}

\section{Data analysis}

\begin{figure}[h!]
\centering 
\includegraphics[width=.45\textwidth,clip]{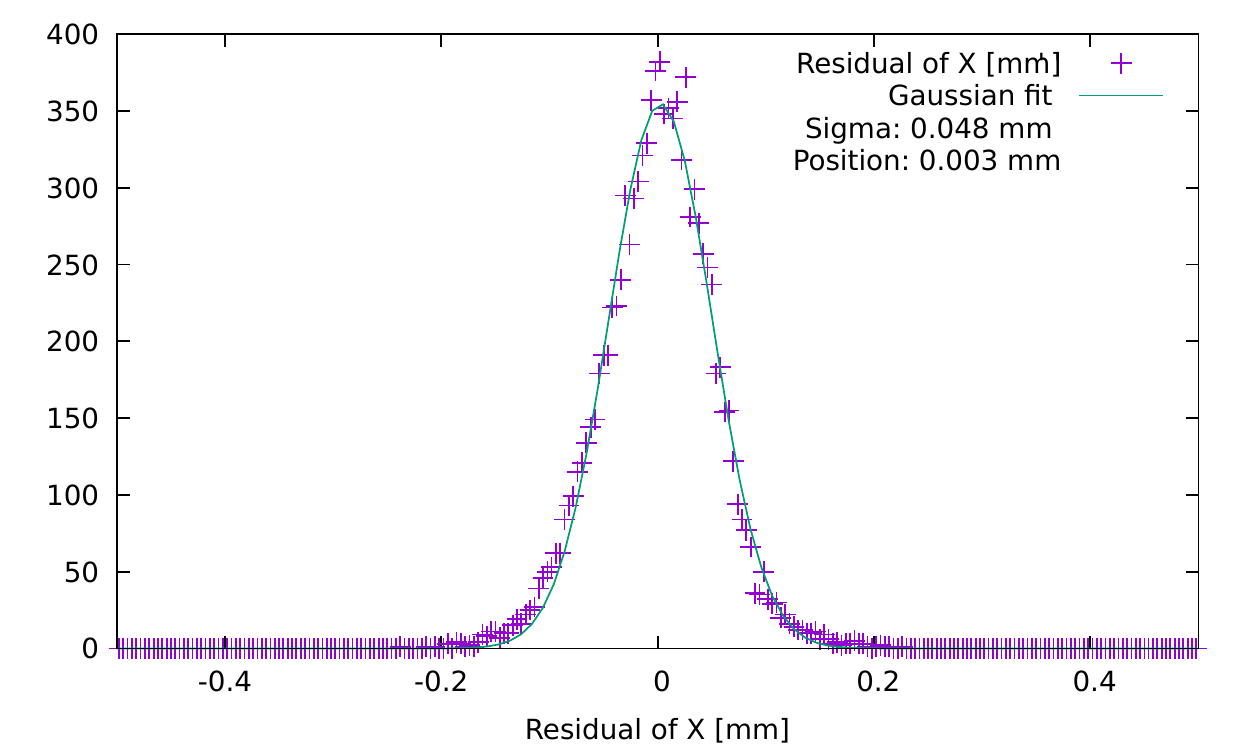}
\qquad
\includegraphics[width=.45\textwidth,origin=c]{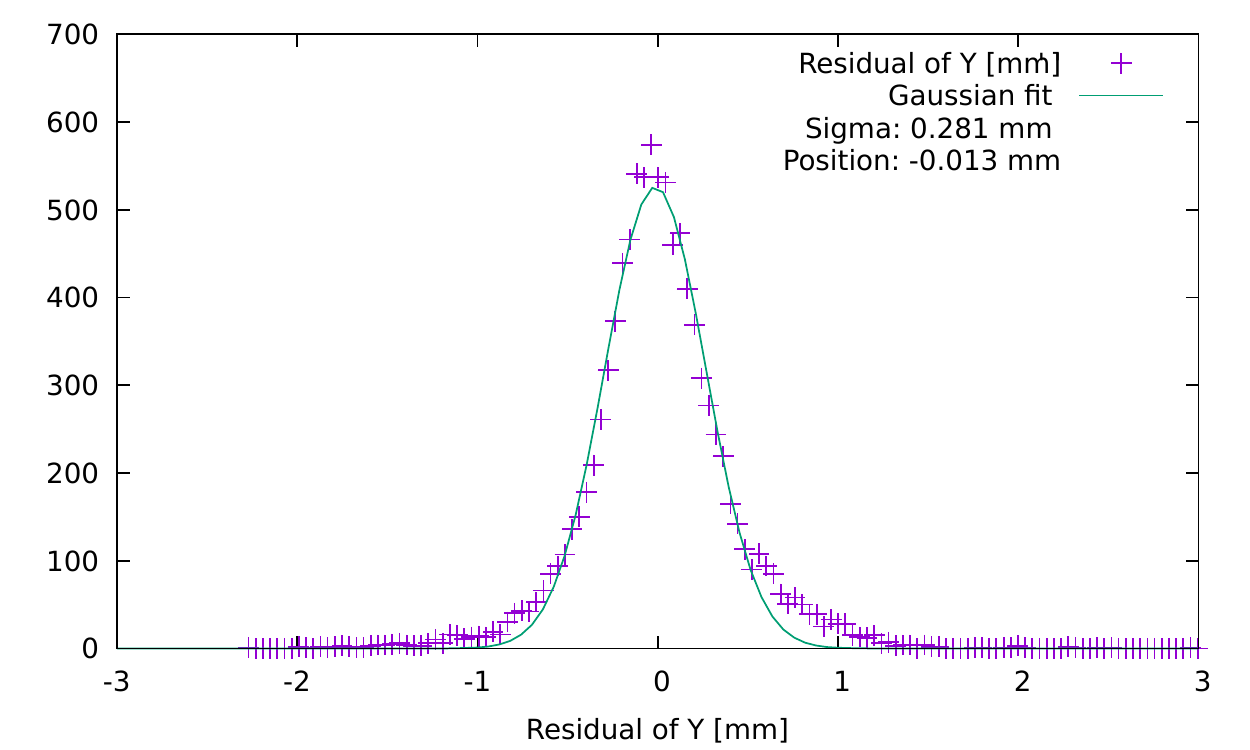}
\caption{\label{fig:residuals} Residuals of 2000 tracks for all pad rows in the direction of the pad rows (left). Residual of the same tracks in the vertical (time) direction (right). A drift velocity of 1.35 cm/$\mu$s was set in this case.}
\end{figure}

In order to process the raw data, a custom C++ analysis code was developed. Since each pad row is read out by its dedicated FPGA board which utilizes its internal clock based on a quartz oscillator, the time stamp of each FPGA is shifted compared to the others. Therefore analysis code matches up events in each file. Afterwards peaks in the signals for each board with a significance of at least 3 $\sigma$ are identified. Flood fill method is used to build clusters around the peaks. Linear regression is applied to all permutations of these cluster in each event. The permutations with a $\chi^2$ per degree of freedom less than one are treated as tracks. In figure \ref{fig:center}. a set of 10000 clusters belonging to tracks are shown projected parellel to the pads (left) and in the vertical (time) direction (right). In figure \ref{fig:residuals}. the difference of the observed cluster positions and the predicted values are plotted. From this, one can derive a spatial resolution of 50 $\mu$m parallel to the pad rows and 280$\mu$m in the vertical direction. In this case the drift velocity was set to 1.35 cm/$\mu$s corresponding to a time resolution of 20 ns. The drift velocity can be set to any practical value in the detector, by adjusting the cathode voltage. The choice of the current drift velocity considers the following aspects: slower drift improves position resolution (as relative time measurement improves due to longer drift time), faster drift reduces off-time background and improves rate tolerance (drift electrons clear faster), whereas practical reduction of cathode voltage simplifies the high voltage system design, preferring slower drift. In our case, the off-time  background was already reduced by the short drift length, so a relatively slow drift, such as 1.35 cm/$\mu$s provided sufficiently good performance.

\newpage

\section{Conclusion and outlook}

The first results of a double-GEM based TPC - designed for ion beam monitoring - are presented. Performance was tested in the ion beam of RIBLL  at the Institute of Modern Physics, in Lanzhou, China. The spatial resolution of the TPC was deretmined to be 50 $\mu$m parallel to the pad rows and 280 $\mu$m in the vertical direction. A second TPC is currently under construction,  to be operated together in tandem configuration in order to to discriminate off-time tracks and to increase angular resolution. 

\section{Acknowledgement}

This work was funded by the bilateral science and technology project between  Hungary and the People's Republic of China, T\'ET 16 CN-1-2016-0008. The authors acknowledge and thankful for the experimental team at IMP, Lanzhou, namely dr. Zheng Yong, dr. Yapeng Zhang, dr. Pengming Zhang and Fengyi Zhao during the beam setup and the measurements.

\end{document}